\documentstyle[prd,aps,twocolumn,draft,psfig,floats]{revtex}

\def\beq{\begin{equation}}
\def\eeq{\end{equation}}
\def\beqa{\begin{eqnarray}}
\def\eeqa{\end{eqnarray}}
\def\bfig{\begin{figure}}
\def\efig{\end{figure}}

\begin{document}
\fnsymbol{footnote}

\wideabs{

\title{Effect of a neutron-star crust on the $r$-mode instability}

\author{Lee Lindblom${}^1$, Benjamin J. Owen${}^2$, and Greg Ushomirsky${}^1$}
\address{${}^1$Theoretical Astrophysics 130-33,
         California Institute of Technology,
         Pasadena, CA 91125}
\address{${}^2$Albert Einstein Institut (Max Planck Institut f\"ur
Gravitationsphysik), Am M\"uhlenberg 1, 14476 Golm bei Potsdam, Germany}
\date{\today}
\maketitle

\begin{abstract}
The presence of a viscous boundary layer under the solid crust of a
neutron star dramatically increases the viscous damping rate of the
fluid $r$-modes.  We improve previous estimates of this damping rate
by including the effect of the Coriolis force on the boundary-layer
eigenfunction and by using more realistic neutron-star models.  If the
crust is assumed to be perfectly rigid, the gravitational radiation
driven instability in the $r$-modes is completely suppressed in
neutron stars colder than about $1.5\times 10^8$~K.  Energy generation
in the boundary layer will heat the star, and will even melt the crust
if the amplitude of the $r$-mode is large enough. We solve the heat
equation explicitly (including the effects of thermal conduction and
neutrino emission) and find that the $r$-mode amplitude needed to melt
the crust is $\alpha_c\approx5\times10^{-3}$ for maximally rotating
neutron stars. If the $r$-mode saturates at an amplitude larger than
$\alpha_c$, the heat generated is sufficient to maintain the outer
layers of the star in a mixed fluid-solid state analogous to the pack
ice on the fringes of the Arctic Ocean.  We argue that in young,
rapidly rotating neutron stars this effect considerably delays the
formation of the crust.  By considering the dissipation in the ice
flow, we show that the final spin frequency of stars with $r$-mode
amplitude of order unity is close to the value estimated for fluid
stars without a crust.
\end{abstract}
}

\narrowtext
\section{Introduction}
\label{sectionI}

The $r$-modes (fluid oscillations governed primarily by the Coriolis
force) have been the focus of considerable attention over the past few
years (see Friedman and Lockitch~\cite{fl99} for a review).  The
gravitational-radiation driven instability of these modes has been
proposed as an explanation for the observed relatively low spin
frequencies of young neutron stars, and of accreting neutron stars in
low-mass x-ray binaries (LMXBs) as well.  The $r$-mode instability may
also provide a source of gravitational waves detectable by the
``enhanced'' LIGO and VIRGO interferometers, the configurations
whose operation is expected to begin in about the year 2006.  In real
neutron stars this instability can only occur when the
gravitational-radiation driving timescale of the $r$-mode is shorter
than the timescales of the various internal dissipation processes that
may occur in the star.  In this paper we re-examine and re-calculate
some of the dissipation timescales associated with the crust of the
neutron star.

Recently Bildsten and Ushomirsky~\cite{BU} made the first estimate of
the effect of a solid crust on the $r$-mode instability.  They found
that the shear dissipation in the viscous boundary layer between the
solid crust and the fluid core decreases the viscous damping timescale
by more than a factor of $10^5$ in old, accreting neutron stars and
more than $10^7$ in hot, young neutron stars. The viscous damping
timescale in these stars is thus comparable to the gravitational
radiation driving timescale, and so Bildsten and Ushomirsky concluded
that the $r$-mode instability is unlikely to play a role in old,
accreting neutron stars.  In hot, young neutron stars they also
predicted that this boundary-layer damping mechanism severely limits
the ability of the $r$-mode instability to reduce the angular momentum
of the star, and hence to produce detectable amounts of gravitational
radiation.

However, the debate over the relevance of the $r$-mode instability to
the observed spin periods of neutron stars is not settled.  Andersson
{\it et al.}~\cite{AJKS} corrected a minor numerical factor in the
work of Bildsten and Ushomirsky~\cite{BU} and used different
neutron-star parameters to obtain a significantly different result for
the critical frequency of the onset of the $r$-mode instability.
Andersson {\it et al.}~\cite{AJKS} estimated the critical frequency
(the frequency of rotation of the star at which the driving and
damping timescales are equal) to be about $40\%$ lower than the
estimate of Bildsten and Ushomirsky~\cite{BU}.  Based on this new
estimate, and contrary to the conclusions of Bildsten and
Ushomirsky~\cite{BU}, Andersson {\it et al.}~\cite{AJKS} inferred that
the $r$-mode instability is likely to be the mechanism that limits the
LMXB spin periods and those of other millisecond pulsars as well.

The calculations of Bildsten and Ushomirsky~\cite{BU} and of Andersson
{\it et al.}~\cite{AJKS} depend on an extremely simple model of the
boundary layer which neglects the Coriolis force, the dominant
restoring force for the $r$-modes.  Rieutord~\cite{R}, building on the
work of Greenspan~\cite{greenspan}, improved this model by finding the
self-consistent solution to the linearized Navier-Stokes equations
(including the Coriolis force terms) throughout the boundary layer.
Rieutord's model of the boundary layer includes the correct angular
structure and results in dissipative timescale estimates that are
significantly shorter than those of Andersson {\it et
al.}~\cite{AJKS}.  Coincidentally, Rieutord's estimates of the
critical spin frequencies agree rather closely with the original
estimates of Bildsten and Ushomirsky~\cite{BU}.  These changes in the
values of the boundary layer dissipation timescale, and the
corresponding changes in the conclusions regarding the physical
relevance of the $r$-mode instability for neutron stars with crusts,
have prompted us to revisit this subject once again.

Our aim is to provide a complete, careful re-derivation of recent
results~\cite{BU,AJKS,R}, including the effects of non-uniform density
stellar models---an important factor neglected by Andersson {\it et
al.}~\cite{AJKS} and Rieutord~\cite{R}.  We use a variety of stellar
models to explore the sensitivity of the results to the poorly-known
neutron-star equation of state.  In summary (see Fig.~\ref{fig1}), we
find that the critical frequency is even greater (by 25\% to 50\%
depending on the equation of state) than that estimated by Bildsten
and Ushomirsky~\cite{BU} and Rieutord~\cite{R}, and double to triple
that estimated by Andersson {\it et al.}~\cite{AJKS}.  Our new results
make it appear unlikely that the $r$-mode instability is responsible
for limiting the spin periods of the LMXBs.

However, the interpretation of our results is somewhat complicated by
the recent work of Levin and Ushomirsky~\cite{LU}. They showed that
mechanical crust-core coupling can reduce the relative velocity
between the crust and the core, thereby reducing the shear and the
viscous dissipaton in the boundary layer.  Let $\Delta v/v$ denote the
difference between the velocities in the inner edge of the crust and
outer edge of the core divided by the velocity of the core.  In the
static, rigid crust case $\Delta v/v=1$, but Levin and
Ushomirsky~\cite{LU} find that this quantity can lie anywhere in the
range $0.05\lesssim \Delta v/v \lesssim 1$.  They find that the
precise value of $\Delta v/v$ varies in a complicated manner with the
spin frequency of the star, and is quite sensitive to the thickness of
the crust and other model parameters.  Throughout this paper we assume
that the crust is static, i.e.\ that $\Delta v/v=1$, but indicate at
appropriate points in the text how to rescale various quantities (such
as dissipation timescales) for $\Delta v/v<1$.  Our calculations
therefore provide only an upper limit on the critical frequency for
the onset of the $r$-mode instability in neutron stars with crusts.
However our limits are easily adjusted once a more realistic value of
$\Delta v/v$ is known.

In young neutron stars, the $r$-mode instability is still a viable
mechanism for spindown even if $\Delta v/v=1$.  In the presence of a
crust, the majority of the viscous dissipation is confined to a very
thin boundary layer.  If the $r$-mode amplitude is larger than some
critical value, the resulting heating in this layer is so intense that
it can compete with neutrino cooling and heat the crust-core interface
to the melting temperature.  This possibility was first suggested by
Owen~\cite{owen}, who crudely estimated the critical amplitude to be
$\alpha_c \approx 10^{-3}$, and the idea of crust re-melting was
suggested again by Andersson {\it et al.}~\cite{AJKS}.  Here we
perform a comprehensive, self-consistent analysis of this heating
effect, including conductive transport of heat into the core and the
crust, and eventual radiation of the excess thermal energy by
neutrinos.  We find that, for maximally rotating neutron stars, the
critical dimensionless $r$-mode amplitude needed to heat the
crust-core interface to the melting temperature at the equator is
$\alpha_c\approx 5\times10^{-3}$.  [This value depends on the spin
frequency and is somewhat larger away from the equator; see
Eq.~(\ref{eq5.13}) and Fig.~\ref{fig4}.]  If the $r$-mode amplitude
grows to a value exceeding $\alpha_c$ in a hot, young neutron star,
the crust will not form as usual and the neutron star will spin down
to a much lower frequency than would have been possible were a crust
present.

What happens if the $r$-mode amplitude does exceed the critical value,
$\alpha_c$?  If the $r$-mode completely melts the crust, the boundary
layer and the heating are removed and the outer layers of the neutron
star quickly cool back down to melting temperature.  If the crust
completely cools and solidifies, the boundary-layer heating due to the
$r$-mode quickly heats the crust back up to melting temperature.  It
is clear that, in the presence of a strong enough $r$-mode, neither a
solid crust nor a pure fluid is possible.  We imagine the situation to
be similar to the pack ice on the Arctic Ocean.  While the $r$-mode
amplitude exceeds the critical value $\alpha_c$ in this picture, the
outer layers will be composed of chunks of crustal ``ice'' floating in
the fluid at the melting temperature of about $10^{10}$K.  The
dissipation mechanism in this pack ice will be a combination of
macroscopic viscosity due to collisions between chunks of ice and
microscopic viscosity due to boundary layers bordering the chunks.
Calculation of the viscosity in this situation would be very
complicated but for the fact that the system must be maintained close
to the melting temperature.  The size of the ice chunks (and other
variables controlling the viscosity) will adjust themselves so that
the heat dissipated in the ice flow balances the neutrino cooling.
This allows us to estimate the $r$-mode damping timescale
quantitatively in the pack ice, without knowing the details of this
complicated process.  We find that, for $r$-mode saturation amplitudes
of order unity, the final spindown frequency is little changed from
that of the purely fluid model of the instability considered over the
past few years.

The rest of this paper is organized as follows. In
Sec.~\ref{sectionII} we re-derive the velocity profile of the
$r$-modes in the boundary layer, and in Sec.~\ref{sectionIII} we
re-derive the energy dissipation in the boundary layer using
techniques and notation that will be more familiar to the relativistic
astrophysics community.  In Sec.~\ref{sectionIV} we apply these
results to the question of the stability of the $r$-modes in hot,
young neutron stars and in older, colder, accreting ones. In
Sec.~\ref{sectionV} we derive the thermal structure of the boundary
layer and find the $r$-mode amplitude $\alpha_c$ necessary to heat the
bottom of the crust to melting. In Sec.~\ref{sectionVI} we argue that
the presence of an $r$-mode with an amplitude greater than $\alpha_c$
will in fact prevent crust formation, and instead lead to the pack-ice
flow described above.  We compute the effective dissipation in this
flow, and consider the implications of the delayed crust formation for
the developent of the $r$-mode instability.  In Sec.~\ref{sectionVII}
we summarize and discuss some of the implications of our results.  In
the Appendix we summarize the relevant thermodynamic properties of the
neutron star matter near the crust-core interface.
 
\section{Structure of the Boundary Layer}
\label{sectionII}

We begin by re-deriving the structure of the $r$-modes in the boundary
layer near a rigid solid crust.  Our analysis improves the initial
studies of Bildsten and Ushomrisky~\cite{BU} and of Andersson {\it et
al.}~\cite{AJKS} by properly evaluating the angular structure of the
boundary layer.  We follow closely the more recent work of
Rieutord~\cite{R}, but employ a notation that is more familiar to the
relativistic astrophysics community and improve his estimates by
allowing non-uniform density stellar models.

Let us decompose the fluid perturbation representing an $r$-mode into
two parts: the eigenfunctions describing the mode in the zero
viscosity limit, $\delta v^a$ and $\delta U\equiv \delta p/\rho$, and
the corrections that must be added to these when viscosity is present,
$\delta\tilde{v}^a$ and $\delta \tilde{U}$.  The velocity correction
$\delta \tilde{v}^a$ must be chosen so that the relative velocity
between the fluid core and the solid crust vanishes: for the case of a
rigid crust this is equivalent to $0=\delta v^a+\delta\tilde{v}^a$.
Thus the viscous corrections to the velocity field are not small, at
least near the crust.  The correction to the hydrodynamic potential
$\delta\tilde{U}$, however, will turn out to be small everywhere.

The equations for the viscous corrections to the velocity field are
obtained by expanding the Navier-Stokes equation to first order in the
amplitude of the perturbation.  We assume that the equilibrium star is
rigidly rotating with angular velocity $\Omega$, and that the temporal
and angular dependence of the mode is $e^{i\omega t+im\varphi}$.
As usual in boundary-layer theory, we assume that the fluid functions
change much more rapidly in the direction perpindicular to the
boundary.  Thus we assume that the angular derivatives are much smaller
than radial derivatives, and neglect them.  Under these assumptions,
the equations that determine the viscous corrections to the fluid
flow are:

\beq
i(\omega+m\Omega)\delta\tilde{v}^r
-2 \Omega r\sin^2\theta\delta\tilde{v}^\varphi =
-\partial_r\delta\tilde{U} 
+ {\case{2}{3}}{\eta\over \rho}\partial_r^2\delta\tilde{v}^r,
\label{eq1.1}
\eeq
\beqa
&&i(\omega+m\Omega)\delta\tilde{v}^\varphi 
+ {2\Omega\over r}\delta \tilde{v}^r
+ 2\Omega \cot\theta \delta\tilde{v}^\theta =\nonumber\\
&&\qquad\qquad\qquad\qquad\qquad\qquad\qquad
{\eta\over\rho}\partial_r^2\bigl(\delta\tilde{v}^\varphi
-{\case{1}{3}}\delta\tilde{v}^r\bigr),\label{eq1.2}
\eeqa
\beq
i(\omega+m\Omega)\delta\tilde{v}^\theta
-2\Omega\cos\theta\sin\theta\delta\tilde{v}^\varphi =
{\eta\over\rho}\partial_r^2\bigl(\delta\tilde{v}^\theta
-{\case{1}{3}}\delta\tilde{v}^r\bigr),\label{eq1.3}
\eeq

\noindent where $\eta$ and $\rho$ are the viscosity and density of
the fluid, respectively.  These equations assume only that the viscous
corrections vary much more rapidly with $r$ than with the angular
coordinates.  Thus we generalize the analysis of Rieutord~\cite{R}: we
do not assume {\it a priori} that the fluid flow is incompressible,
or that the equilibrium is of uniform density.

The solutions to these equations depend exponentially on $r$ with a
characteristic length scale $d$:

\beq
d^2 = {\eta\over 2\Omega\rho}\biggl|_{r=R_c}.\label{eq1.4}
\eeq

\noindent The radial velocity correction $\delta\tilde{v}^r$ must
vanish both at the boundary and deep within the fluid.  Given the
exponential nature of the solutions, it follows that
$\delta\tilde{v}^r=0$ everywhere within the fluid.  With this
simplification, Eqs.~(\ref{eq1.2})--(\ref{eq1.3}) determine the
velocity corrections $\delta\tilde{v}^\theta$ and
$\delta\tilde{v}^\varphi$, while Eq.~(\ref{eq1.1}) determines
$\delta\tilde{U}$ in terms of $\delta\tilde{v}^\varphi$.
Eqs.~(\ref{eq1.2}) and (\ref{eq1.3}) may be re-written in the
following form:

\beq
\Bigl[d^2\partial_r^2-i\Bigl({\sigma\over 2\Omega}+\cos\theta\Bigr)\Bigr]
\bigl(\delta\tilde{v}^\theta
+i\sin\theta\delta\tilde{v}^\varphi\bigr)=0,\label{eq1.5}
\eeq
\beq
\Bigl[d^2\partial_r^2-i\Bigl({\sigma\over 2\Omega}-\cos\theta\Bigr)\Bigr]
\bigl(\delta\tilde{v}^\theta
-i\sin\theta\delta\tilde{v}^\varphi\bigr)=0,\label{eq1.6}
\eeq

\noindent where $\sigma=\omega+m\Omega$ is the mode frequency in the
rotating frame.

It is straightforward now to write down the general solutions to these
equations, and then to impose the boundary condition, $0=\delta
v^a+\delta\tilde{v}^a$, at the inner edge of the crust $r=R_c$.  These
solutions are given by

\beq
\delta\tilde{v}^\theta = - \delta v^\theta(R_c) \Lambda_+(r,\theta)
       -i \sin\theta\, \delta v^\varphi(R_c)\Lambda_-(r,\theta),
\label{eq1.7}
\eeq

\beq
\sin\theta\,\delta\tilde{v}^\varphi 
= - \sin\theta\,\delta v^\varphi(R_c) \Lambda_+(r,\theta)
       +i \delta v^\theta(R_c)\Lambda_-(r,\theta),\label{eq1.8} 
\eeq

\noindent where $\delta v^a(R_c)$ is the standard non-viscous $r$-mode 
velocity perturbation, evaluated at the radius of the inner edge of the
crust $r=R_c$.  The functions $\Lambda_\pm(r,\theta)$ are defined
by

\beq
\Lambda_\pm(r,\theta) = 
\case{1}{2} e^{-\zeta\sqrt{i(\cos\theta+{\sigma/2\Omega})}}
\pm\case{1}{2} e^{-\zeta\sqrt{i(\cos\theta-{\sigma/ 2\Omega})}},
\label{eq1.9}
\eeq

\noindent where $\zeta$ is the dimensionless radial parameter

\beq
\zeta = {R_c -r\over d},\label{eq1.10}
\eeq

\noindent and the characteristic thickness of the boundary layer, $d$,
is defined in Eq.~(\ref{eq1.4}). For $r$-modes the frequency of the
mode (as measured in the co-rotating frame of the fluid) that appears
in the definition of $\Lambda_\pm$ has the value
$\sigma/2\Omega=1/(m+1)$.  For rapidly rotating neutron stars
$d\approx( 10^{8}$K$/T)\,\,$cm, where $T$ is the temperature in the
boundary layer.  Therefore, $d/R_c\ll 1$.

\section{Dissipation in the Boundary Layer}

\label{sectionIII}

The shear of the $r$-mode velocity field is dominated by the rapid
radial change in $\delta \tilde{v}^a$ through the boundary layer.
Thus, up to terms of order $d/R_c$, the square of the shear tensor in
the boundary layer is given by

\beq
\delta\sigma_{ab}^*\delta\sigma^{ab} = \case{1}{2} R^2_c
\Bigl( |\partial_r\delta \tilde{v}^\theta|^2
+\sin^2\theta\,|\partial_r\delta \tilde{v}^\varphi|^2\Bigr).\label{eq2.1}
\eeq

\noindent The angular structure of $\delta\tilde{v}^a$ in
Eqs.~(\ref{eq1.7})--(\ref{eq1.8}) is determined in part by the
structure of the dissipation-free velocity field $\delta v^a$.  For
convenience we may write the non-dissipative $r$-mode velocity field as

\beq
\delta v^\theta = -iAr^{m-1}\sin^{m-1}\theta \,e^{i(\omega t + m\varphi)},
\label{eq2.2}
\eeq

\beq
\delta v^\varphi = Ar^{m-1}{\sin^{m-2}\theta} \,\cos\theta\, 
e^{i(\omega t + m\varphi)},\label{eq2.3}
\eeq

\noindent where $A$ is a normalization constant.  It is straightforward
then to evaluate the square of the shear tensor:

\beq
\delta\sigma_{ab}^*\delta\sigma^{ab} = {|A|^2r^{2m}\over 8\,d^2}F(r,\theta),
\label{eq2.4}
\eeq

\noindent where 
\beqa
F(r,\theta)=\sin^{2m-2}\theta\,\bigl[&&(1-\cos\theta)^2p_+^2e^{-\zeta p_+}
\nonumber\\&&+(1+\cos\theta)^2p_-^2e^{-\zeta p_-}\bigr],
\label{eq2.41}
\eeqa

\noindent and

\beq
p_\pm = \sqrt{2|\cos\theta \pm 1/(m+1)|}.\label{eq2.5}
\eeq

Now integrate the energy dissipation rate due to shear viscosity over
the fluid interior to the crust, ignoring terms of order $d/R_c$:

\beq
{d\tilde{E}\over dt}
=-\int 2\eta\delta\sigma_{ab}^*\delta\sigma^{ab}d^{\,3}x
=2\sqrt{2}\pi |A|^2 \Omega d R_c^{2m+2}\rho_c {\cal I}_m,\label{eq2.6}
\eeq

\noindent where ${\cal I}_m$ is defined by

\beq
{\cal I}_m = \int_0^\pi \sin^{2m-1}\theta(1+\cos\theta)^2
\sqrt{|\cos\theta-1/(m+1)|}d\theta.\label{eq2.7}
\eeq

\noindent For the case of primary interest to us, $m=2$, this
integral has the value ${\cal I}_2 = 0.80411$~\cite{R}.

Now we can define the viscous timescale for dissipation in the
boundary layer:

\beq
{1\over \tau_v} = - {1\over 2\tilde{E}}{d\tilde{E}\over dt}.
\label{eq2.8}
\eeq

\noindent Using the expression for $d\tilde{E}/dt$ derived above, and
the usual expression for the energy $\tilde{E}$ we find:

\beqa
\tau_v &&= {1\over 2\Omega}{2^{m+3/2}(m+1)!\over m(2m+1)!!{\cal I}_m}
\times\nonumber\\
&&\qquad\qquad\sqrt{2\Omega R_c^2\rho_c\over\eta_c}
\int_0^{R_c} {\rho\over\rho_c}\left({r\over R_c}\right)^{2m+2} {dr\over R_c}.
\label{eq2.9}
\eeqa

\noindent Here the quantities $R_c$, $\rho_c$ and $\eta_c$ are the
radius, density, and the viscosity of the fluid at the outer edge of
the core.  We note that, while the viscous dissipation rate in the
boundary layer, Eq.~(\ref{eq2.6}), depends only on quantities local to
the boundary layer ($d$, $\rho_c$, $R_c$, $\Omega$), the timescale
$\tau_v$ depends also on the global structure of the mode.  This is
due to the fact that, while most of the energy dissipation takes place
in the boundary layer, most of the energy in the mode is not localized
there.  For realistic neutron stars this expression,
Eq.~(\ref{eq2.9}), for the viscous timescale is about a factor of two
larger than the one obtained by Rieutord~\cite{R}, who assumed a
uniform-density stellar model.

In deriving our expression for the viscous boundary layer dissipation
timescale, Eq.~(\ref{eq2.9}), we assumed that the crust is rigid and
hence static in the rotating frame. The motion of the crust due to the
mechanical coupling to the core \cite{LU} effectively increases
$\tau_v$ by a factor of $(\Delta v/v)^{-2}$.

\section{Stability of the $r$-Modes}
\label{sectionIV}

In this section we evaluate viscous timescales for neutron stars
(based on slowly rotating Newtonian models) constructed from a set of
``realistic'' equations of state, and use these timescales to evaluate
the stability of these stars.  In neutron stars colder than about
$10^9$~K the shear viscosity is expected to be dominated by
electron-electron scattering.  The viscosity associated with this
process is given by \cite{flowers-itoh,CL}

\beq\label{eta-ee}
\eta^{ee}=6.0\times 10^6 \rho^2 T^{-2},\label{eq3.1}
\eeq

\noindent where all quantities are given in cgs units, and $T$ is
measured in K.  For temperatures above about $10^{9}$K,
neutron-neutron scattering provides the dominant dissipation
mechanism.  In this range the viscosity is given by
\cite{flowers-itoh,CL}

\beq\label{eta-nn}
\eta^{nn}=347\rho^{9/4} T^{-2}.\label{eq3.11}
\eeq

We find it useful to factor the angular velocity and temperature
dependence from the viscous timescale defined in Eq.~(\ref{eq2.9}).
Thus we define a fiducial viscous timescale $\tilde{\tau}_v$
such that

\beq
\tau_v = \tilde{\tau}_v \left({\Omega_o\over\Omega}\right)^{1/2}
\left({T\over
10^8{\rm K}}\right),\label{eq3.2}
\eeq

\noindent where $\Omega_o=\sqrt{\pi G \bar{\rho}}$.  We have evaluated
these fiducial viscous timescales (for each type of viscous
dissipation) for 1.4$M_\odot$ neutron star models based on a variety
of realistic equations of state~\cite{paris} as well as the standard
$n=1$ polytrope with a radius of 12.53~km.  These results are
summarized in Table~\ref{table1}, along with other relevant properties
of these stellar models.  In particular we also include the total
radius $R$, the radius of the fluid core $R_c$, the energy contained
in an excited $r$-mode
$\tilde{e}=\tilde{E}(\Omega_o/\alpha\Omega)^2$~ergs for the case of a
fully fluid star, the energy contained in an excited $r$-mode
$\tilde{e}_c$ in the case with rigid crust, the fiducial gravitational
radiation timescale $\tilde{\tau}_{GR}$, and a critical temperature
$T_c$ (defined below).  We note that the $\tau_v$ values listed in
Table~\ref{table1} presume that the crust is rigid and does not move
in the corotating frame.  To take into account the motion of the
crust, multiply $\tilde\tau_v^{ee}$ and $\tilde\tau_v^{nn}$ by
$(\Delta v/v)^{-2}$~\cite{note1}.

\begin{table}
\caption{Properties of 1.4$M_\odot$ neutron stars with rigid crusts
for densities below $\rho_c= 1.5\times 10^{14}$g/cm${}^3$. Times are
given in seconds, lengths in kilometers, temperatures in units of
$10^8$K, and energies in units of $10^{51}$~ergs.}
\label{table1}
\begin{tabular}{lr@{.}lr@{.}lr@{.}lr@{.}lr@{.}lr@{.}lr@{.}lr@{.}lr@{.}r}
EOS\tablenote[1]{The various equations of state used here are described 
in Bonazzola, Frieben and Gourgoulhon~\protect\cite{paris}.}
&\multicolumn{2}{c}{$R$}
&\multicolumn{2}{c}{$R_c$}
&\multicolumn{2}{c}{$\tilde e_c$}
&\multicolumn{2}{c}{${\tilde e}$}
&\multicolumn{2}{c}{$\tilde{\tau}_{GR}$}
&\multicolumn{2}{c}{$\tilde{\tau}_v^{ee}$}
&\multicolumn{2}{c}{$T_c^{ee}$}
&\multicolumn{2}{c}{$\tilde{\tau}_v^{nn}$}
&\multicolumn{2}{c}{$T_c^{nn}$}\\
\hline
n1poly& 12&53& 11&01&  1&94& 2&53&  4&25& 23&3& 1&69& 52&0& 0&76\\
BJI   & 15&04& 12&16&  1&30& 1&72& 13&11& 24&6& 4&96& 54&8& 2&23\\
DiazII& 15&80& 12&60&  1&20& 1&67& 17&37& 24&3& 6&64& 54&2& 2&98\\
FP    & 13&18& 11&34&  1&93& 2&18&  5&24& 25&9& 1&88& 57&8& 0&84\\
Glend1& 16&47& 12&74&  0&97& 1&53& 25&32& 23&4& 10&1& 52&2& 4&51\\
Glend2& 16&84& 12&95&  0&96& 1&53& 28&12& 23&7& 11&0& 52&8& 4&95\\
HKP   & 16&43& 13&27&  1&32& 1&76& 18&41& 24&6& 6&97& 54&8& 3&13\\
PandN & 12&77& 10&85&  1&84& 2&08&  4&84& 26&9& 1&67& 59&9& 0&75\\
SHW   & 14&92& 12&89&  1&97& 2&23&  8&40& 25&0& 3&11& 55&8& 1&40\\
WFF3  & 12&98& 11&21&  2&10& 2&33&  4&52& 27&7& 1&52& 61&8& 0&68\\
WGW   & 14&64& 12&38&  1&85& 2&17&  8&32& 26&8& 2&88& 59&8& 1&29
\end{tabular}
\end{table}

The gravitational radiation timescale is evaluated here using the
formalism developed by Lindblom, Owen, and Morsink~\cite{LOM}.  This
timescale is given by

\beqa
{1\over\tau_{GR}} = &&{32\pi G \Omega^{2m+2}\over c^{2m+3}}
{(m-1)^{2m}\over [(2m+1)!!]^2}\nonumber\\
&&\qquad\times\left({m+2\over m+1}\right)^{2m+2}
\int_0^{R_c}\rho r^{2m+2} dr.\label{eq3.3}
\eeqa

\noindent Since we presume that the crust is static in the rotating
frame, the integral in Eq.~(\ref{eq3.3}) extends only over the
interior of the fluid core, rather than throughout the star.
Consequently, these gravitational radiation timescales are somewhat
longer (typically about $30\%$ longer) than those computed earlier for
hotter, completely fluid stars~\cite{note1}. The fiducial
gravitational radiation timescale $\tilde{\tau}_{GR}$ given in
Table~\ref{table1} is defined by

\beq
\tau_{GR} = \tilde{\tau}_{GR} \left({\Omega_o\over\Omega}\right)^{2m+2}.
\label{eq3.4}
\eeq

Gravitational radiation tends to drive the $r$-modes unstable, while
viscosity suppresses the instability.  We define the critical angular
velocity $\Omega_c$, above which the $r$-mode is unstable, by the
condition $\tau_v=\tau_{GR}$:

\beq
{\Omega_c\over\Omega_o} 
= \left({\tilde{\tau}_{GR}\over\tilde{\tau}_v}\right)^{2/11}
\left({10^8{\rm K}\over T}\right)^{2/11}.\label{eq3.5}
\eeq

\begin{figure}
\centerline{\psfig{file=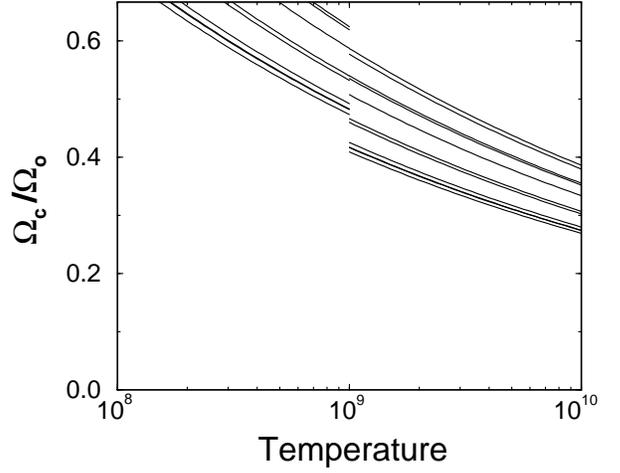,height=2.4in}}\vskip 0.3cm
\caption{Temperature dependence of the critical angular velocities of
$1.4 M_\odot$ neutron stars constructed with a number of realistic
equations of state.}
\label{fig1}
\end{figure}

\noindent Figure~\ref{fig1} illustrates the temperature dependence of
the critical angular velocity for $1.4M_\odot$ neutron stars
constructed from a variety of realistic equations of state.  The
discontinuities in these curves at $T=10^9\,$K occur because the
superfluid transition changes the viscosity from electron-electron
scattering at low temperatures to neutron-neutron scattering. The
outer edge of the core in these models is taken to be
$\rho_c=1.5\times 10^{14}$~g~cm${}^{-3}$~\cite{peth-raven-lorenz}.
Equation~(\ref{eq3.5}) and the curves in Fig.~\ref{fig1} neglect the motion
of the crust in the rotating frame, i.e., they assume that $\Delta
v/v=1$.  In general, $\Omega_c$ is a factor of $(\Delta v/v)^{4/11}$
smaller than that given in Eq.~(\ref{eq3.5}).  Thus, $\Omega_c$
displayed in Fig.~\ref{fig1} provides an upper bound on the critical
frequency for realistic neutron stars.  In these expressions for
$\Omega_c$ we neglect the effects of bulk viscosity, which are
unimportant for $T\lesssim 10^{10}$~K.

The angular velocity of a neutron star can never exceed some maximum
value $\Omega_{\rm max}\approx \case{2}{3}\Omega_o$~\cite{paris}.
Thus, there is a critical temperature below which the gravitational
radiation instability is completely suppressed by viscosity.  This
critical temperature is given by

\beq
{T_c\over 10^8{\rm K}}=\left({\Omega_o\over\Omega_{\rm max}}\right)^{11/2}
{\tilde{\tau}_{GR}\over\tilde{\tau}_v}
\approx \left({3\over 2}\right)^{11/2}{\tilde{\tau}_{GR}\over\tilde{\tau}_v}.
\label{eq3.6}
\eeq

\noindent In terms of $T_c$ then the critical angular velocity can
be expressed in a particularly simple form:

\beq
{\Omega_c\over\Omega_o}={\Omega_{\rm max}\over\Omega_o}
\left({T_c\over T}\right)^{2/11}\approx \frac{2}{3}
\left({T_c\over T}\right)^{2/11}. 
\label{eq3.7}
\eeq

\begin{figure}
\centerline{\psfig{file=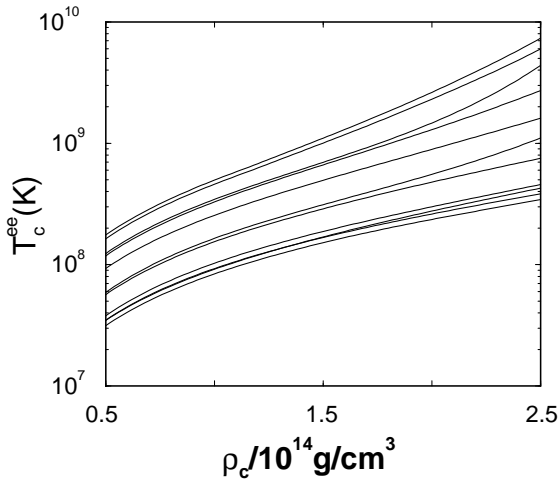,height=2.4in}}
\vskip 0.3cm
\caption{Critical temperature $T_c$ as a function of the crust
formation density $\rho_c$ for a number of realistic equations of
state.}
\label{fig2}
\end{figure}

\noindent The values of this critical temperature are given in
Table~\ref{table1} for the case $\rho_c=1.5\times 10^{14}$~g~cm${}^{-3}$.
Since the exact density $\rho_c$ where crust formation begins is only
poorly known, we explore in Fig.~\ref{fig2} the dependence of $T_c$
on $\rho_c$.  The motion of the crust would change $T_c$ of
Eq.~(\ref{eq3.6}) by the factor $(\Delta v/v)^2$, and so the curves in
Fig.~\ref{fig2} provide an upper bound on $T_c$ for realistic
neutron star models.

\section{Thermal Structure of the Boundary Layer}
\label{sectionV}

Viscous dissipation in the boundary layer between the core and the
crust deposits thermal energy into a very thin layer of material.
Previous authors assumed that this heat is effectively conducted away
from the boundary layer, so that the star remains essentially
isothermal.  Clearly, this situation is idealized, and viscous
dissipation will raise the temperature at the crust-core interface.
The viscous heating competes with thermal conduction away from the
boundary layer, as well as neutrino emission from the crust and the
core. In this section we evaluate the $r$-mode amplitude needed to
raise the temperature at the boundary layer to a given value.  Our
particular interest is to determine the minimum $r$-mode amplitude
needed to raise the temperature to
$T_m\approx10^{10}(\rho/\rho_c)^{1/3}$~K, the melting temperature of
the crust (see the Appendix for a discussion of the microphysics
employed in this paper).  If this occurs, the $r$-modes may continue
to be unstable to much lower angular velocities (see
Sec.~\ref{sectionVI}).

In the discussion that follows, it will be necessary to know the
explicit temperature dependences of the thermal conductivity and
neutrino emissivity.  As described in the Appendix, these
temperature dependences are

\beq
\kappa={\tilde{\kappa}\over T},\label{eq5.3}
\eeq

\beq
\epsilon=\tilde{\epsilon} T_{10}^n,\label{eq5.4}
\eeq

\noindent where $T_{10}=T/10^{10}$~K, and the exponent in the
emissivity equation is $n=8$ in the fluid core and $n=6$ in the crust.
The prefactors $\tilde{\kappa}$ and $\tilde{\epsilon}$ are independent
of temperature and are given by $\tilde{\kappa}_<=1.5\times10^{31}$,
$\tilde{\epsilon}_<=8.6\times10^{28}$ in the core and
$\tilde{\kappa}_>=2.8\times10^{30}$,
$\tilde{\epsilon}_>=1.5\times10^{25}$ in the crust.  The quantities
$\kappa$ and $\epsilon$ are in cgs units.  Note that for notational
convenience in subsequent computations, the temperature in
Eq.~(\ref{eq5.3}) is measured in K, while in Eq.~(\ref{eq5.4}) it is
measured in $10^{10}$~K.

Let us first neglect the thermal conductivity altogether, and ask what
$r$-mode amplitude is necessary to heat the boundary layer to $T_m$ if
the heat is radiated exclusively by neutrino emission from the
boundary layer itself. The rate $d\tilde E/dt$ at which the shear
deposits energy into the vicinity of the boundary layer is given by
Eq.~(\ref{eq2.6}), where the amplitude $A$ is related to the
dimensionless $r$-mode amplitude, as defined by Lindblom, Owen and
Morsink~\cite{LOM}, via

\beq
\alpha=\sqrt{16\pi\over 5}{R\over \Omega}A.\label{eq5.7}
\eeq

\noindent We equate this rate to the neutrino emission rate in the
boundary layer, $4\pi R_c^2 d \tilde\epsilon_{<} T_{10}^8$, to obtain
$\alpha_c({\rm local})$, the $r$-mode amplitude necessary to keep the
crust-core interface at the melting temperature if thermal conduction
is not important:

\beqa \alpha_c({\rm local})\, &&= \frac{R}{R_c}
\left(\frac{16\pi\sqrt{2}\,\tilde\epsilon_< T_{m,10}^8}
{5{\cal I}_2\rho_cR_c^2\Omega^{3}} \right)^{1/2} 
\nonumber \\ && = 1.3\times10^{-4}
\left(\frac{\Omega_o}{\Omega}\right)^{3/2} T_{m,10}^4.
\label{alpha-local}
\eeqa

\noindent In this equation the numerical prefactor has been evaluated
using the standard $n=1$ polytropic stellar model. This estimate sets
the lower bound on the critical melting amplitude, since heat
conduction increases the volume that radiates heat by neutrino
emission, and hence more viscous dissipation is required to maintain
the interface at a certain temperature.

We now show how to evaluate the effect of finite conductivity.  The
general equation for the thermal evolution of the material in a
neutron star is

\beq
C_v\partial_tT = -\epsilon + \vec{\nabla}\cdot(\kappa\vec{\nabla}T)
+2\eta\delta\sigma^*_{ab}\sigma^{ab},\label{eq5.1}
\eeq

\noindent where $T$ is the temperature, $C_v$ is the specific heat at
constant density, $\epsilon$ is the neutrino emissivity, and $\kappa$
is the thermal conductivity.  This is a time-dependent equation, and
in general the overall cooling of the star due to Urca neutrino
emission must be followed along with the heating in the boundary
layer.  Let $\ell$ denote the thickness of the region adjacent to the
boundary layer whose temperature is raised above the ambient by the
viscous dissipation.  On a timescale $t_{\rm diff}\approx
C_v\ell^2/\kappa$ the heat generated in the boundary layer will
diffuse throughout this larger region.  In the neutron-star matter
near the boundary layer this timescale is $t_{\rm diff}\approx
8\times10^{-5}(\ell/d)^2$~s, which depends on the temperature only
through the boundary layer thickness, $d\propto T^{-1}$.  The neutron
star as a whole cools on the Urca cooling timescale, $t_{\rm
cool}\approx30\ T_{10}^{-6}$~s~\cite{ST}.  If the diffusion timescale
is shorter than the cooling timescale, then the temperature
distribution in the region near the boundary layer will estabilish a
quasi-equilibrium state in which the excess heat is conducted into a
large enough volume for it to be radiated away by neutrinos.  The
temperature at the inner edge of this region slowly decreases,
tracking the overall cooling of the star.  As we shall see below, the
width of this quasi-equilibrium layer is a few hundred times the
thickness of the boundary layer itself.  Thus $t_{\rm diff} < t_{\rm
cool}$ and so a quasi-equilibrium state exists in which $C_v\partial_t
T$ can be neglected compared to the heat conduction term
$\partial_r(\kappa\partial_r T)$.

In the quasi-equilibrium region, but outside the boundary layer, the
temperature distribution is described approximately by

\beq
\partial_r(\kappa\partial_r T)=\epsilon.
\label{eq5.2}
\eeq

\noindent Using the simple dependence of $\kappa$ and $\epsilon$ on
temperature, given by Eqs.~(\ref{eq5.3}) and~(\ref{eq5.4}), it is
straightforward to obtain a first integral of Eq.~(\ref{eq5.2}),

\beq
\left({\partial T_{10}\over\partial r}\right)^2=
{2\tilde{\epsilon}\over
n\tilde{\kappa}}T_{10}^2(T_{10}^n-T_{o,10}^n),\label{eq5.5}
\eeq

\noindent where $T_o$ is the ambient temperature in the core where the
heat flux $\kappa\partial_r T$ tends to zero.

The viscous energy generation in the boundary layer is determined by
the shear given by Eq.~(\ref{eq2.4}).  Here we re-express this
energy generation rate as

\beq
2\eta\delta\sigma^*_{ab}\delta\sigma^{ab}
=\alpha^2{5\rho_cR_c^{\,4}\Omega_o^3\over 32\pi R^2}
\left({\Omega\over\Omega_o}\right)^3F(r,\theta),\label{eq5.6}
\eeq

\noindent where the function $F(r,\theta)$ is defined in
Eq.~(\ref{eq2.41}).  This function falls off exponentially away from
the boundary layer with the length scale $d$.  Thus $F$ can be
reasonably approximated as a delta function,

\beq
F(r,\theta)\approx
 \sqrt{{4\eta\over 3\Omega\rho_c}}
{\delta(r-R_c)\over f^2(\theta)},\label{eq5.8}
\eeq

\noindent where the angular function $f(\theta)$ is defined by

\beq
f^{-2}(\theta)=\sqrt{\case{3}{8}}\sin^2\theta\Bigl[(1-\cos\theta)^2p_+
+(1+\cos\theta)^2p_-\Bigr].
\label{eq5.9}
\eeq

Now we return to the full equation for the thermal distribution in
the vicinity of the boundary layer:

\beq
\partial_r(\kappa\partial_r T)=\epsilon-
2\eta\delta\sigma^*_{ab}\delta\sigma^{ab}.
\label{eq5.10}
\eeq

\noindent This equation can be integrated analytically, using
Eq.~(\ref{eq5.5}), when we approximate the heating in the viscous
boundary layer by Eq.~(\ref{eq5.8}).  The $r$-mode amplitude $\alpha_c$
needed to raise the temperature to the value $T_m$ is

\beqa
\alpha_c^2 {5R_c^4\sqrt{\eta\rho_c \Omega^{5}}
\over 16\sqrt{3}\pi R^2f^2(\theta)}=
&&\sqrt{\case{1}{4}\tilde{\epsilon}_<\tilde{\kappa}_<(T_{m,10}^8-T_{o,10}^8)}
\nonumber \\
&&+\sqrt{\case{1}{3}\tilde{\epsilon}_>\tilde{\kappa}_>(T_{m,10}^6-T_{o,10}^6)}.
\label{eq5.11}
\eeqa

\noindent For the values of the microphysical parameters described
above, the critical amplitude satisfies

\beqa
{\alpha_c^2\over f^2(\theta)T_{m,10}^5} 
\left({\Omega\over\Omega_o}\right)^{5/2}&&=
8.0\times10^{-6}\sqrt{1-{T_o^8\over T_m^8}}\nonumber\\
&&+{5.3\times10^{-8}\over T_{m,10}}\sqrt{1-{T_o^6\over T_m^6}}.
\label{eq5.12}
\eeqa

\begin{figure}
\centerline{\psfig{file=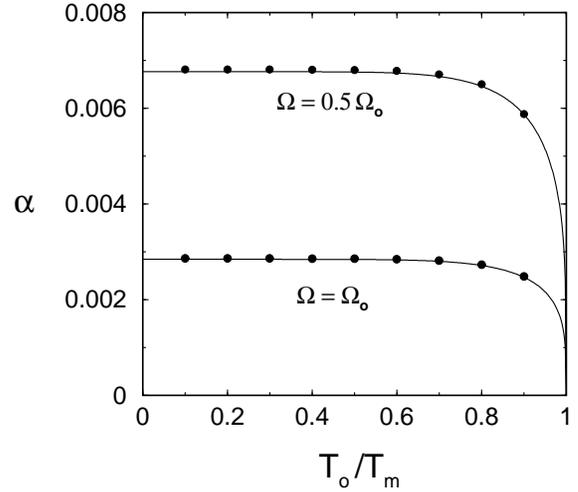,height=2.5in}}\vskip 0.3cm
\caption{Dependence of the critical amplitude $\alpha$ on $T_o$ for
stars with $\Omega=0.5\Omega_o$. The smooth curve is the analytical
expression given in Eq.~(\ref{eq5.11}), while the points are numerical
solutions to the heat equation without the delta-function
approximation for the boundary layer heating term.}
\label{fig3}
\end{figure}

We illustrate in Fig.~\ref{fig3} the dependence of this critical
amplitude on the ambient temperature $T_o$ for the case
$T_m=10^{10}$~K and $\theta=\pi/2$.  The solid curve is the analytical
expression given in Eq.~(\ref{eq5.12}).  For comparison, we also
include a selection of points computed numerically by solving the full
differential equation, Eq.~(\ref{eq5.10}), without making the
assumption that the heat source is a delta function.  Clearly, the
analytical approximation is extremely good.  We also see that, because
of the steep dependence of the neutrino cooling rate on the
temperature, the value of $\alpha_c$ is rather insensitive to $T_o$,
except when $T_o\approx T_m$.

The prefactor of the second term on the right side of
Eq.~(\ref{eq5.12}) is much smaller than the prefactor of the first
term.  Thus we see that the core plays a much more important role in
determining $\alpha_c$ than the crust.  This is easy to understand: At
$T=T_m$, the neutrino emissivity in the core is approximately $5000$
times larger than in the crust.  Since we demand that the crust-core
interface is kept at a fixed temperature, much more heat is emitted on
the core side of the interface than on the crust side. Therefore, the
heat flux into the core must be much higher than the flux into the
crust.  Since the heat flux into the crust can be neglected
(i.e. $\tilde{\epsilon}_<\tilde{\kappa}_< \gg
\tilde{\epsilon}_>\tilde{\kappa}_>$), we may obtain a much simpler
expression for the critical amplitude:

\beq
\alpha_c\approx2.8\times10^{-3}T_{m,10}^{5/2}f(\theta)
\left({\Omega_o\over\Omega}\right)^{5/4}.\label{eq5.13}
\eeq

\noindent for the case $T_o\ll T_m$.  For a maximally rotating neutron
star, with $\Omega_{\rm max}\approx \case{2}{3}\Omega_o$, the
prefactor in the above equation is $4.7\times10^{-3}$.

\begin{figure}
\centerline{\psfig{file=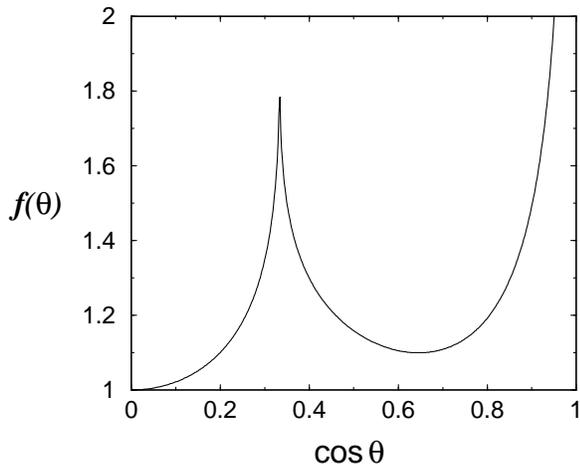,height=2.4in}}\vskip 0.3cm
\caption{Function $f$ that determines the angular dependence
of the critical amplitude $\alpha$.}
\label{fig4}
\end{figure}

The function $f(\theta)$, illustrated in Fig.~\ref{fig4}, determines
the angular dependence of the critical melting amplitude.  Melting the
crust near the poles requires a higher $r$-mode amplitude than melting
at the equator. However, over a wide range of angles the critical
amplitude needed to heat the interface to a given temperature is the
same as the equatorial ($\theta=\pi/2$) value to within a factor of
two.

The critical melting amplitude $\alpha_c$ is roughly 20 times
$\alpha_c({\rm local})$, our estimate that neglects the effect of
thermal conduction, Eq.~(\ref{alpha-local}).  From this we may deduce
that the thickness of the layer into which thermal energy is conducted
before being radiated away is of order
$\ell/d\approx\alpha_c^2/\alpha_c^2({\rm local})\approx 400$. From our
numerical solutions, we confirm that the thermal flux in the core
falls from its value at the crust-core interface to half that value at
approximately $500 d$.  Thus the thickness of the layer that radiates
the excess thermal energy is small enough (just a few centimeters) to
justify the quasi-equilibrium analysis which we used.

Finally, we note that the heating rate used in this calculation
neglects the motion of the crust in the rotating frame~\cite{LU}.  The
motion of the crust increases the critical melting amplitude by the
factor of $(\Delta v/v)^{-1}$.

\section{Dissipation in the Ice Flow}
\label{sectionVI}

As a young neutron star cools, its temperature quickly falls below the
freezing temperature and a crust begins to form.  If no unstable
$r$-mode is present, a solid crust forms as usual.  However if the
star is rapidly rotating, the $m=2$ $r$-mode will be driven unstable,
and may grow to an appreciable amplitude even before the crust begins
to form.  Therefore, the crust would have to form in the presence of
the shearing motion of the $r$-mode.  Even in the presence of the
crust, the $r$-mode is unstable at high enough frequencies, as
illustrated in Fig.~\ref{fig1}.  Once the amplitude of the mode grows
beyond the critical value given in Eq.~(\ref{eq5.13}), the dissipation
in the boundary layer would be sufficient to re-melt the crust if it
formed.  If the crust fails to form or is completely melted away, the
boundary layer heating would dissappear and the dissipation 
would not be sufficient to prevent freezing.  Clearly, the state of
the outer layers of the star can be neither pure solid nor pure fluid.
We imagine that instead a mixed state will form: a neutron-star ice
flow, much like the pack ice on the surface of the Arctic Ocean.

Microscopic viscosity will be replaced in this ice flow by dissipation
that is dominated by collisions between macroscopic chunks of crust,
the boundary layers around the ice chunks, or other mechanisms.
However, regardless of the detailed dissipation mechanism, this flow
is self-regulating: as long as the amplitude of the $r$-mode remains
above the critical value of Eq.~(\ref{eq5.13}), the dissipation must,
on average, be enough to keep the crust at around the melting
temperature.  Any less dissipation, and the crust would completely
freeze, thus giving rise to a boundary layer which would re-melt the
crust.  Any more dissipation and the crust would simply melt, then
promptly re-form as the fluid viscosity alone is insufficient to keep
the outer layers hot.  The assumption of self-regulating flow allows
us to compute the dissipation rate {\it regardless\/} of the details
of the dissipation mechanism.

The dominant energy sink at high temperatures is bremsstrahlung
neutrino emission, with emissivity given by Eq.~(\ref{crust-bremss}).
Thus the viscous dissipation in the ice flow must just balance the
losses due to neutrino emission.  Presuming that the outer layers of
the neutron star are kept at the local melting temperature given by
Eq.~(\ref{Tmelt}), we can determine the total energy dissipated in the
ice flow, and hence an effective timescale $\tau_{\rm ice}$ on which
this ice flow dissipation will damp the $r$-mode:

\beq {1\over\tau_{\rm ice}}={1\over 2E}\int
\tilde{\epsilon}\,T_{m,10}^6\, d^{3}x.
\label{eq6.1}
\eeq

\noindent Displaying explicitly the scalings of $\tau_{\rm ice}$ with
the mode amplitude $\alpha$ and the angular velocity $\Omega$, we find

\beq \tau_{\rm ice} =  \alpha^2 \left(\Omega \over
\Omega_o \right)^2\tilde\tau_{\rm ice}.  \label{eq6.3}\eeq

\noindent The fiducial timescales $\tilde\tau_{\rm ice}$ range from
$1.4\times 10^8$~s to $1.3\times 10^9$~s for the ``realistic''
equations of state discussed above.  Unlike the other dissipative
timescales in this problem, this one depends on the mode amplitude
$\alpha$.

The $r$-mode will continue to be unstable, and thus will continue to
spin down the neutron star by emitting gravitational radiation, as
long as the gravitational radiation timescale is shorter than the ice
flow viscosity timescale, and its amplitude remains above the value
needed to sustain the ice flow.  The critical angular velocity
where the equality is achieved and the mode becomes stable is

\beq {\Omega_c\over\Omega_o}= \left({\tilde\tau_{GR} \over
\tilde\tau_{\rm ice}} \right)^{1/8} \alpha^{-1/4} .
\label{eq6.2}
\eeq

\noindent For $\alpha=1$, this critical angular velocity has the value
$0.093\Omega_o$ for the standard $n=1$ polytropic stellar model, and
values that range from $0.086\Omega_o$ to $0.14\Omega_o$ for the
realistic models considered here.  (Note that the gravitational
radiation timescale $\tau_{GR}$ used here is the one appropriate for
a mode that extends to the surface of the star, not the one given in
Table~\ref{table1} for a mode that extends only to the edge of the
core.)  Thus we see that if the amplitude of the $r$-mode saturates at
a value that is close to unity, the instability will continue---even
in the presence of a crust---until the angular velocity of the star is
a small fraction of $\Omega_o$.

Is the ice-flow picture described above consistent?  Clearly the
dissipation due to molecular viscosity is insufficient to keep the
outer layers of the neutron star hot, but can collisions between
macroscopic ice chunks accomplish this?  Let us estimate the
characteristic size of the ice chunks needed to produce the needed
macroscopic viscosity.  We assume that the ice chunks are of average
size $D$, and the mean free path between them is $\lambda$.  If
$\lambda\gg D$, then the collisions between the chunks are rare, and
do not significantly raise the viscosity compared to the purely fluid
value (see \S~22 of Landau and Lifshitz~\cite{LL-fluids}). We
therefore consider the regime where the ice chunks are densely packed,
with $D\gtrsim\lambda$, and behave like a fluid.  We follow the
discussion of Haff~\cite{haff} and Borderies {\it et al.}~\cite{BGT}.

The ice chunks mainly follow the $r$-mode velocity flow, but due to
the collisions between themselves acquire a random component of
velocity, $\tilde v$.  The viscosity in such a flow is given by
$\eta_{\rm ice}\approx \rho D^2 \tilde v/\lambda$ \cite{haff,BGT}, and
is a factor of $D/\lambda$ bigger than the usual molecular viscosity.
This is because each particle transports momentum across a distance
$D$ while only having traveled the (possibly much smaller) distance
$\lambda$ between each collision.

The collisions between the chunks are inelastic with a fraction
$\gamma$ of the collision energy per chunk, $\rho D^3 \tilde v^2$,
being converted to heat.  Considerations of icy particle collisions in
the temperature regime appropriate for Saturn's rings put
$\gamma\approx0.7-0.9$ (see Borderies {\it et al.}~\cite{BGT-unsolved}
for a review).  Since our flow is near the melting temperature, we
expect very inelastic collisions as well, i.e., $\gamma\approx1$.  The
collision rate is just $\tilde v/\lambda$, so the energy dissipation
per unit volume in the flow is $\gamma\rho\tilde v^3/\lambda$.  For
the $r$-modes the shear dissipation rate is $2\eta_{\rm ice}
\delta\sigma^*_{ab}\delta\sigma^{ab}\approx \eta_{\rm
ice}\alpha^2\Omega^2$, so $\tilde v\approx\gamma^{-1/2}\alpha\Omega
D$.  Thus when the collisions are inelastic, the random component of
the ice chunk velocity is comparable to the velocity difference
between the neighboring chunks in the overall ice flow.  The viscosity
of this granular flow is therefore given by

\begin{eqnarray}
 \eta_{\rm ice} \approx
\frac{\rho\alpha\Omega D^3}{\gamma^{1/2}\lambda}
\approx 10^{18} \
\frac{\alpha D^3}{\gamma^{1/2}\lambda}
\left(\frac{\rho}{\rho_c}\right)
\left(\frac{\Omega}{\Omega_o}\right), 
\end{eqnarray}

\noindent where $\lambda$ and $D$ are measured in centimeters and
$\eta_{\rm ice}$ is in cgs units.  Not surprisingly, this viscosity is
much larger than the microscopic viscosities given by
Eqs.~(\ref{eta-ee}) and~(\ref{eta-nn}).  

The energy dissipation rate in the ice flow is obtained by integrating
$\eta_{\rm ice} \delta\sigma^*_{ab}\delta\sigma^{ab}$ over the outer
layers of the star:

\beq \frac{dE}{dt} \approx
 \frac{(\alpha\Omega D)^3 M_{\rm cr}}{\gamma^{1/2}\lambda},
\eeq

\noindent where $M_{\rm cr}$ is the mass of the material at densities
less than $\rho_c$.  Using the $r$-mode energy for an $n=1$ polytrope
(see Table~\ref{table1}), the damping time due to this form
of viscous dissipation is

\beq \tau_{\rm ice}\approx 10^8 {\rm~s~}
\frac{\gamma^{1/2}\lambda}{\alpha D^3}
\left(\frac{\Omega_o}{\Omega}\right) \left(\frac{0.05 M}{M_{\rm
cr}}\right) \eeq

\noindent The heat deposited by the ice flow into the star is radiated
by neutrinos on the timescale evaluated in Eq.~(\ref{eq6.3}).
Equating these two timescales gives an estimate of the average ice
chunk size necessary to keep the crust at the local melting
temperature:

\begin{equation}
D \approx 1{\rm~cm~}
\left[\left(\frac{10^8{\rm~s}}{\tilde\tau_{\rm ice}}\right)
\left(\frac{\gamma^{1/2}\lambda}{D}\right)
\left(\frac{\Omega_o}{\alpha\Omega}\right)^{3}
\left(\frac{0.05 M}{M_{\rm cr}}\right)\right]^{1/2}.\nonumber
\end{equation}

\noindent Our estimate of the ice chunk size depends only weakly on
the unknown inelasticity of the collisions and the ratio $\lambda/D$.
We do not expect $\lambda$ to be much smaller than $D$, because if it
were, the ice chunks would probably lock together into bigger pieces,
leading to increased friction and dissipation.  Moreover, as argued
above, we expect $\gamma\approx1$ as well.  Thus for typical values of
$\tilde\tau_{\rm ice}$ evaluated above, the value of the ice chunk
size needed to keep the outer layers of the star at the melting
temperature is $D\approx 1\,$cm.

Several conclusions can be drawn from this estimate. First, the
smaller the mode amplitude, the larger the ice chunks have to be in
order to provide enough friction (recall that the viscosity is
proportional to $D^2$) to keep the fluid at the melting temperature in
the face of neutrino losses.  The chunk size exceeds the radius of the
star when $\alpha\lesssim10^{-4} (\Omega_o/\Omega)(\lambda/D)^{1/3}$.
Hence, if the mode amplitude is smaller than this value, the ice-flow
mechanism is unable to keep the outer part of the star in a melted
state, and it must freeze.  Since this amplitude is smaller than the
critical melting amplitude computed in the previous section, this does
not further restrict the range of mode amplitudes where the ice flow
is expected to occur.  Second, even for $\alpha\approx1$ the chunk
size is small but not microscopic.  So our picture of the ice flow is
consistent throughout the regime of interest.

The argument that the ice flow maintains the star's outer layers near
melting depends on the ability of the $r$-mode to re-heat a solid
crust back to the verge of melting, should a crust form through some
fluctuation.  Up to this point we assumed that this heating is
instantaneous once $\alpha$ exceeds $\alpha_c$.  This is not an issue
if $\Omega>\Omega_c$ [the critical angular velocity for the $r$-mode
instability, Eq.~(\ref{eq3.5})], since the amplitude of the $r$-mode
is growing due to radiation reaction in this case.  However, if
$\Omega<\Omega_c$, the $r$-mode is decaying due to the dissipation in
the boundary layer even while it is re-heating the crust.  Can the
$r$-mode re-heat the crust and re-form the ice flow before being
damped?  A simple comparison of the energy needed to melt the crust,
$E_m$, to the mode energy, $\tilde{E}$ (see Table~\ref{table1}), is
inadequate, since it neglects the continuing energy input from
gravitational radiation reaction.

The $r$-mode melts the crust on a timescale

\beq 
\tau_m = \frac{E_m}{\beta (d\tilde{E}/dt)_v}=
\tau_v\frac{E_m}{2\beta\tilde{E}}, 
\eeq

\noindent where the factor $\beta$ is the fraction of heat flowing
into the crust, which is given approximately by
$\beta=(4\tilde\epsilon_>\tilde\kappa_>/3\tilde
\epsilon_<\tilde\kappa_<)^{1/2}\approx0.007$ 
(see Eq.~[\ref{eq5.5}]).  While it is melting the crust, the $r$-mode
is also damped at the rate $1/\tau_d = 1/\tau_v - 1/\tau_{GR}$.  The
damping rate is exactly zero on the stability curve $\Omega=\Omega_c$.
However, since the $r$-mode has a finite amplitude, it will continue
to spin the star down below the stability line while melting the
crust.

Let $\Delta\Omega$ be the change in spin frequency, due to the
$r$-mode evolution, during the time $\tau_m$.  For small $\alpha$, the
spindown rate is given by Eq.~(3.14) of Owen {\it et
al.}~\cite{olcsva98}:

\beq
\frac{1}{\Omega}\frac{d\Omega}{dt} \approx \frac{0.2\alpha^2}{\tau_v}.
\eeq

\noindent Thus, in the time it takes to melt the crust, the star spins
down by $\Delta\Omega /\Omega_c \approx 0.1\alpha^2
E_m/\beta\tilde{E}$ (which is independent of $\alpha$ since $\tilde
E\propto\alpha^2$).  By using a Taylor expansion, it is easy to show
that the $r$-mode damping time for $\Omega=\Omega_c-\Delta\Omega$ is
approximately

\beq 
\tau_d\approx \frac{2}{11}\frac{\Omega_c}{\Delta\Omega}\tau_v=
\frac{2\beta\tilde E}{\alpha^2 E_m}\tau_v.
\eeq

\noindent If $\tau_m< \tau_d$, we conclude that the $r$-mode can re-melt 
the crust and so re-create the ice flow before being damped out.  This
condition places the following limit on the mode amplitude $\alpha$,

\beq
\alpha \gtrsim \frac{E_m}{2\beta\tilde{e}}
\left(\frac{\Omega_o}{\Omega}\right)^2 \approx
3\times10^{-3}\ \left(\frac{\Omega_o}{\Omega}\right)^2.
\label{survive}
\eeq

\noindent In the above equation, we assumed that the energy required
to re-melt the crust is of order $kT$ per nucleus (since the crust is
near the melting temperature), i.e.\ $E_m\approx10^{47} T_{m,10}$~erg.
This amplitude is very close, both in the absolute magnitude and in
the scaling with $\Omega$ and $T_m$, to the critical melting amplitude
$\alpha_c$. Hence, so long as the $r$-mode amplitude is large enough
to raise the crust-core interface to the melting temperature, the
energy contained in the mode is enough to melt the crust even if the
$r$-mode is no longer linearly unstable.  Since the
criterion~(\ref{survive}) does not depend on $\Delta v/v$ while
$\alpha_c$ (Eq.~[\ref{eq5.13}]) increases with $\Delta v/v$, the
conclusion does not change even when the motion of the crust is taken
into account.

\section{Discussion}
\label{sectionVII}

We have computed stability curves (critical frequency as a function of
temperature) for the onset of the $r$-mode instability in a neutron
star with a (laminar) viscous boundary layer under a solid crust. We
improve previous calculations by including the effect of the Coriolis
force (the dominant restoring force) on the boundary layer and by
using realistic neutron-star models---two important ingredients which
have not been combined in previous work. Our stability results are
summarized in Eqs.~(\ref{eq3.6}) and~(\ref{eq3.7}),
Table~\ref{table1}, and Figs.~\ref{fig1} and~\ref{fig2}.  If the
neutron star crust is rigid and does not move in the rotating frame,
then our results imply that the $r$-modes are not unstable in any of
the accreting neutron stars in LMXBs or millisecond pulsars.  However,
if the relative velocity amplitude between the core and the crust,
$\Delta v/v$ is significantly smaller than 1, as recent calculations
for constant-density stars suggest~\cite{LU}, then our results
constitute only the upper limits on the critical frequencies and
temperatures for the onset of the $r$-mode instability.  A
self-consistent calculation of the $r$-mode eigenfunctions in the
presence of a realistic crust is necessary to settle (at least for
unmagnetized neutron stars) the question of linear stability of the
$r$-modes.

We find that localized heating in the boundary layer between the solid
crust and the fluid core can successfully compete both with the heat
conduction away from the boundary layer and with neutrino emission.
In Section~\ref{sectionV}, we computed the critical $r$-mode
amplitude, Eq.~(\ref{eq5.13}), needed to raise the temperature of the
boundary layer to the crust melting temperature while the interior of
the star far away from the boundary layer remains at a much lower
temperature.  The amplitude required for this is rather small, of
order $10^{-2}$, and is comparable to that necessary to break up the
crust via purely mechanical coupling.

Based on the smallness of the melting amplitude, we argue in
Section~\ref{sectionVI} that the spindown scenarios for neutron stars
with crusts advanced by previous authors may need to be significantly
modified.  So long as the $r$-mode amplitude remains above the melting
value, a completely solid crust is not possible.  Instead, the outer
layers of a neutron star will resemble an ice pack which flows along
with the $r$-mode motion.  Viscosity in the ice pack adjusts itself to
maintain the outer layers of the neutron star near the crust's melting
temperature, regardless of the details of the dissipation in the ice
flow.  The ice flow can persist even if the $r$-mode is no longer
linearly unstable according to the instability criterion for neutron
stars with crusts, since an $r$-mode with an amplitude greater than
the minimum value given in Eq.~(\ref{survive}) can melt the crust
before being damped by boundary-layer friction.  Therefore, if the
crust is melted either during spindown of young neutron stars, or in
the final stage of a thermal runaway in LMXBs, the final spin
frequency is not set by the boundary-layer damping time, but instead
by the balance between $\tau_{\rm ice}$ and $\tau_{\rm GR}$.  This
leads to much smaller final spin frequencies than computed by previous
authors.  A neutron star with an $r$-mode amplitude large enough to
create the ice pack can spin down to a frequency nearly as low as a
purely fluid star [see Eq.~(\ref{eq6.2})], provided the saturation
amplitude of a purely fluid $r$-mode of is order unity.

Very recently, Wu {\it et al.}~\cite{wu} evaluated the effect of
a turbulent boundary layer between the crust and the core on the
stability of the $r$-modes.  They find that the turbulent dissipation
timescale depends on the $r$-mode amplitude and therefore determines a
saturation amplitude, which they estimate (in our notation) as
$\alpha_{\rm sat} \approx 1.5\times10^{-2} (\Omega/\Omega_o)^{5}
(\Delta v/v)^{-3}$, or about $2.0\times10^{-3}$ for a maximally
rotating star with a rigid crust.  Inserting Eq.~(10) of Wu {\it et
al.}~\cite{wu} into our Eq.~(\ref{eq5.10}), we find that a turbulent
boundary layer changes the critical melting amplitude at the equator
to

\beq
\alpha_c \approx 5.6\times10^{-4} \left(\Omega_o \over \Omega\right)
\left(\Delta v\over v\right)^{-1} ,
\eeq
 
\noindent or $\alpha_c \approx 7.9\times10^{-4}$ for a maximally
rotating polytropic neutron star with a rigid crust.  Comparing with
Eq.~(\ref{eq5.13}), we conclude that the crust can be heated to the
melting temperature at even smaller amplitudes than in the presence of
a laminar boundary layer.  Thus, if the $r$-mode amplitude in a
newborn neutron star exceeds the critical value given above when the
temperature drops below the crust's melting temperature, then
crust formation will be delayed (as we argued in
Sec.~\ref{sectionVI}).  In this case the star will spin down to
frequencies close to those predicted for crustless fluid stars,
provided the saturation amplitude of the $r$-mode is of order unity.
\vfill\eject

\acknowledgments

We thank C. Cutler, P. Goldreich, Y. Levin, and B. Schutz for
helpful discussions concerning this work.  This work has been
supported by NSF grants PHY--9796079 and AST--9618537, and by NASA
grants NAG5--4093 and NGC5--7034.

\appendix

\section*{Thermodynamic Properties of Neutron-Star Matter}

The strength of electrostatic interactions between the nuclei in
neutron-star matter is typically expressed in terms of the Coulomb
coupling parameter,

\beq
\Gamma = \frac{Z^2 e^2}{a kT},
\eeq

\noindent where $Z$ is the nuclear charge, and $a=(3 A_{\rm cell}
m_b/4\pi\rho)^{1/3}$ is the radius of the Wigner-Seitz cell with
$A_{\rm cell}$ nucleons in it.  According to the recent
molecular-dynamics simulations of Farouki and Hamaguchi~\cite{FH},
such a classical one-component plasma crystallizes when $\Gamma\gtrsim
173$. However, the details of the nucleon-nucleon interaction that
determine $Z$ and $A_{\rm cell}$ near the bottom of the crust,
$\rho=\rho_c$, are still not completely understood.  Depending on the
particular interaction model, values of $Z$ ranging from approximately
$30$ to $50$, and $A_{\rm cell}$ in the range of $500$ to $1000$ are
obtained.  These values result in melting temperatures within a factor
of two of $10^{10}$~K. (See Pethick and Ravenhall~\cite{PR} for a
review of the nuclear physics, Douchin and Haensel~\cite{Sly-crust}
for the latest calculation, and Haensel~\cite{haensel-melting} for a
discussion of melting temperatures appropriate for young neutron stars
and LMXBs.)  Moreover, for some nuclear force models, nuclei at
$\rho\gtrsim10^{14}$~g~cm$^{-3}$ may resemble rods, plates, and tubes,
rather than spheres (Lorentz {\it et al.}~\cite{LRP}).  Melting
temperatures of such exotic phases of matter have not been calculated,
but it is reasonable to assume~\cite{PP} that they will be comparable
to those of ``ordinary'' matter with spherical nuclei.  In the light
of these uncertainties, we adopt

\beq\label{Tmelt}
T_m=10^{10} \left(\frac{\rho}{\rho_c}\right)^{1/3}{\rm~K},
\eeq

\noindent as the fiducial melting temperature of crustal matter.

The thermal conductivity in the neutron-star core is dominated by
electron-electron collisions.  A convenient fit to the conductivity is
given by Flowers and Itoh~\cite{fl-itoh-fits}:

\beq
\kappa=\frac{1.5\times10^{21}}{T_{10}}
{\rm~erg~cm}^{-1}{\rm~s}^{-1}{\rm~K}^{-1},
\eeq

\noindent at $\rho=\rho_c$, where $T_{10}=T/10^{10}$~K.  The neutrino
emissivity in the core is given by the usual modified Urca formula
\cite{Urca1,Urca2}: 

\beq
\epsilon=8.6\times10^{28}\, T_{10}^8{\rm~erg~cm}^{-3}{\rm~s}^{-1},
\eeq

\noindent at $\rho=\rho_c$. These lead to the expressions for the core
values, $\tilde{\kappa}_<=1.5\times 10^{31}$ and
$\tilde{\epsilon}_<=8.6\times 10^{28}$, used in
Sec.~\ref{sectionV}.

The microphysics in the crust is, in general, quite a bit more
complicated. However, we are primarily interested in the conditions at
or near the melting temperature.  For $T\gtrsim2\times10^{10}$~K, the
crustal conductivity is also dominated by electron-electron
collisions, and even at $T\approx10^{10}$~K, to within a factor of
two,

\beq
\kappa\approx\kappa_{ee}\approx
\frac{2.8\times10^{20}}{T_{10}}{\rm~erg~cm}^{-1}{\rm~s}^{-1}{\rm~K}^{-1},
\eeq

\noindent where we approximated the fits of Flowers and
Itoh~\cite{fl-itoh-fits} for the temperature regime of interest.  For
temperatures lower than this, the conductivity is dominated by
electron-phonon scattering, which is approximately constant,
$\kappa_{\rm e-ph}\approx 10^{20}$ for temperatures in excess of the
Debye temperature, $T_d\approx 5\times 10^9 (\rho/\rho_c)^{1/2}$~K.
For $T\ll T_d$, the electron-phonon scattering has the same
temperature dependence (but a different prefactor) as
electron-electron scattering, $\kappa_{\rm e-ph}\approx 1\times
10^{19}/T_{10}$.  We also find that the recent calculations of
neutrino-pair bremsstrahlung in the crust (Haensel {\it et
al.}~\cite{bremss1} and Kaminker {\it et al.}~\cite{bremss2}) can be
reasonably approximated by 

\beq\label{crust-bremss}
\epsilon\approx1.5\times10^{25}\left(\frac{\rho}{\rho_c}\right)\,
T_{10}^6{\rm~erg~cm}^{-3}{\rm~s}^{-1}.
\eeq

\noindent These expressions lead to the crust values,
$\tilde{\kappa}_>=2.8\times 10^{30}$ and $\tilde{\epsilon}_>=1.5\times
10^{25}$ that we use in Sec.~\ref{sectionV}.  Since the crust does not
play a significant role in determining the temperature at the boundary
layer (as shown in Sec.~\ref{sectionV}), including a detailed
treatment of the microphysics there is not essential.


\begin{references}
\bibitem{note1} For a star with a non-rigid crust the mode energy
$\tilde{e}_c$ is slightly greater and the gravitational radiation
timescale $\tau_{GR}$ is slightly shorter than the values displayed in
Table~\ref{table1}.  However the differences depend on $\Delta v/v$ in
a manner more complicated than simple scaling, and therefore we do not
explore them in this paper.
\end{references}
\end{document}